\documentclass[fleqn,twoside]{article}
\usepackage{espcrc2}

\usepackage{amssymb,amsfonts,mathrsfs}

\newcommand{\AmS}{{\protect\the\textfont2
  A\kern-.1667em\lower.5ex\hbox{M}\kern-.125emS}}

\title{Renormalization Group Flow in Algebraic Holography}

\author{Pedro Lauridsen Ribeiro\address[FMA]{Departamento de F\'\i sica
	Matem\'atica -- Instituto de F\'\i sica, Universidade de
	S\~ao Paulo \\ 
        CP 66.318 \quad 05315-970 \quad S\~ao Paulo, SP -- Brasil \\
	Email: \texttt{pribeiro@fma.if.usp.br}}
        \thanks{This project is supported by FAPESP under grant no. 
	01/14360-1.}}
       
\begin{document}

\begin{abstract}
An approach to the Holographic Renormalization Group in the context
of Rehren duality -- a structural form of the AdS-CFT correspondence, 
in the context of Local Quantum Physics (Algebraic QFT) -- is
proposed. Special attention to the issue of UV/IR connection is paid.
\vspace{1pc}
\end{abstract}

\maketitle

\section{Introduction}

The remarkable scaling properties of the AdS-CFT
correspondence\cite{malda} have been proven extremely useful as a
calculational device for the scaling behaviour of holographic pairs,
known as the \emph{Holographic Renormalization Group (RG)}. 
In this communication, the aforementioned scaling properties will be
studied in a somewhat different context. We'll adopt the formalism 
of Local Quantum Physics\cite{araki}, in terms 
of C*-algebras of local observables, for which a structural 
(i.e., model-independent) form 
of the AdS-CFT correspondence has been proven by Rehren. 
This form, called \emph{Algebraic Holography} or \emph{Rehren 
duality}, is, however, a statement for local observables in a 
\emph{fixed} spacetime background: some aspects that are linked 
to unrestricted general covariance, such as 
conformal anomalies\cite{sken}, are quite obscure in our setting. 
Nevertheless, it will be shown that other phenomena, such as the 
UV/IR connection\cite{suss1} and the very duality between the 
dynamics between different leaves of the Poincar\'e foliation of 
anti-de Sitter (AdS) spacetime and the RG flow of the boundary 
(Minkowski) theory, here appear naturally.

\section{Algebraic Holography}

Here we shall review the main ideas underlying Rehren duality.

\subsection{Geometry}

Let the $d$-dimensional Minkowski space be denoted by $\mathscr{M}_d$,
$d\geq 2$, and its conformal compactification by $c\mathscr{M}_d$. 
Anti-de Sitter space-time with $d+1$ dimensions (denoted 
$\mathit{AdS}_{d+1}$, or simply $\mathit{AdS}$, when no confusion
about dimensionality arises) can be described by its embedding in
$\mathbb{R}^{d+2}$, given by the quadric (let $R\,=\,1$)
\begin{eqnarray}\label{ads}
X^0X^0-\mathbf{X\cdot X}-X^dX^d+X^{d+1}X^{d+1}\;=\;R^2 \nonumber\\
(\mathbf{X}\;=\;(X^1,X^2,\ldots,X^{d-1})).
\end{eqnarray}
Its boundary at spatial infinity is timelike, and conformal to 
$c\mathscr{M}_d$. 
The (identity component of the) isometry group of $AdS_{d+1}$ 
is $SO_e(d,2)$ ($\mathit{AdS}$ group). A crucial feature of it is that it 
coincides with the conformal group of $\mathscr{M}_d$. 
In $\mathit{AdS}_{d+1}$, we can define a causally complete region 
\begin{eqnarray}\label{wedge}
\mathscr{W}_0 & = & \{X\in AdS_{d+1}\;: \nonumber\\
 & & X^d+X^{d+1}-\sqrt{1+\mathbf{X}\cdot\mathbf{X}}>|X^0|\},
\end{eqnarray} 
called a (standard) \emph{wedge}. The class of all wedges is obtained
by the action of the $\mathit{AdS}$ group on $\mathscr{W}_0$. 
Using a Poincar\'e coordinate patch $(z,x^\mu):z>0,\, 
x^\mu\in\mathscr{M}_d,\,\mu=0,\ldots,d-1$
\begin{equation}\label{poinc}
\left\{ \begin{array}{l@{\;=\;}l}
X^\mu & \frac{1}{z}x^\mu \\
X^d & \frac{1-z^2}{2z}+\frac{1}{2z}x_\mu x^\mu \\
X^{d+1} & \frac{1+z^2}{2z}-\frac{1}{2z}x_\mu x^\mu
\end{array} \right. 
\end{equation} in the region $\{X\in AdS_{d+1}:X^d+X^{d+1}>0\}$, 
we can write the $AdS$ metric as
\begin{equation}\label{m1}
ds^2\,=\,\frac{1}{z^2}(dx_\mu dx^\mu-dz^2)
\end{equation}
and the standard wedge as
\begin{equation}
\mathscr{W}_0\;:=\;\{(x^\mu,z)\;:\;
\sqrt{z^2+\mathbf{x}\cdot\mathbf{x}}<1-|x^0|\}. 
\end{equation}
By taking the
limit $z\;\rightarrow\;0$, we reach the boundary
$\mathscr{M}_d$, yielding the intersection of $\mathscr{W}_0$ with it:
\begin{equation}
\mathscr{K}_0\;=\;\alpha(\mathscr{W}_0)\;=\;\{x^\mu\in\mathscr{M}_d\;:\;
|\mathbf{x}|<1-|x^0|\}.
\label{cone}
\end{equation} 
This is the (standard) \emph{diamond} in Minkowski space. 
Let $\alpha$ be a bijection from the set of all wedges in $AdS_{d+1}$ to
their intersections with $AdS_{d+1}$ boundary (namely, the conformal
class of diamonds), such that:

\begin{itemize}
\item $\alpha(\mathscr{W}_0)\,=\,\mathscr{K}_0$, and
\item It interwines both actions of $SO_e(d,2)$ on $AdS_{d+1}$ and $c\mathscr{M}_d$.
\end{itemize}

Such an $\alpha$ preserves inclusions and causal
complements. Moreover, wedges and the conformal class of diamonds
generate all open sets in their respective spaces.

\subsection{Observables}

We work with theories of local observables, defined as a
(isotonous, anti-de Sitter covariant and causal) net of unital 
C*-algebras $\mathfrak{A}(\mathscr{O})$ indexed by all 
regions $\mathscr{O}$ in $AdS$\cite{rehren}. 
Notice that we mean \emph{causal} with respect to the \emph{covering} of
$AdS$, since pure $AdS$ possesses closed timelike curves. 
The correspondence $\alpha$ built above between the set of wedges
$\mathfrak{W}\,=\,\{\Lambda\mathscr{W}_0\subset AdS_{d+1}:
\Lambda\in SO(d,2)\}$ and the conformal class of diamonds 
$\mathfrak{K}\,=\,\{\alpha(\Lambda \mathscr{W}_0)\subset
\mathscr{M}_d:\Lambda\in SO(d,2)\}$ allows us to build a 
(isotonous, conformally covariant and causal) net of unital
C*-algebras $\mathfrak{B}(\mathscr{O})$ over the regions $\mathscr{O}
\subset\mathscr{M}_d$ by defining

\begin{equation}
\mathfrak{B}(\alpha(\mathscr{W}))\,:=\,\mathfrak{A}(\mathscr{W}),\,
\mathscr{W}\in\mathfrak{W}.
\end{equation}

Therefore, we can state the

\textsc{Theorem (Rehren Duality)\rm\cite{rehren}.}\quad\emph{From an anti-de
Sitter covariant and local net of observables in
$AdS_{d+1}$, one can build an one-to-one correspondence to a
conformally covariant and local net of observables in
$c\mathscr{M}_d$, interwining the action of $SO_e(d,2)$ and preserving
inclusions and causal complements (net isomorphism).}

\section{Leaf nets and holographic RG flow}

Consider again the Poincar\'e coordinate patch (\ref{poinc}), which can be
seen as to define a foliation (warped product) of the region 
$AdS^+_{d+1}\,=\,\{X\in AdS_{d+1}:X^d+X^{d+1}>0\}$. For each fixed
$z$, the (timelike) \emph{leaf} $(z,x^\mu)$ is conformal to Minkowski 
spacetime by a conformal factor of $1/z^2\,=\,(X^d+X^{d+1})^2$.
Two other crucial property of these leaves\footnote{Also called 
\emph{branes} in \cite{bros1}, after the work of Randall and 
Sundrum\cite{randall}. Here, however, we find such terminology 
inadequate, since, in our case, such ``branes'' are not considered 
as being dynamical objects; thus, we'll stick to the name ``leaf'' 
throughout the entire text.} are:

\begin{enumerate}
\item The subgroup of $SO_e(d,2)$ defined by
\begin{eqnarray}\label{lorentz}
(z,x^\mu)\mapsto(z,\Lambda^\mu_\nu x^\nu+a^\mu),\nonumber\\
\Lambda\in SO(d,1),\,a^\mu\in\mathscr{M}_d
\end{eqnarray} (Poincar\'e subgroup) leaves every leaf invariant and
acts as the ($d$-dimensional) Poincar\'e group on each leaf, 
including the boundary $(z=0)$; 
\item The subgroup of $SO_e(d,2)$ defined by
\begin{equation}\label{dilation}
(z,x^\mu)\mapsto(\lambda z,\lambda x^\mu),\,\lambda>0
\end{equation} (dilation subgroup) leaves the region $AdS^+_{d+1}$
invariant, and acts as the dilation subgroup of $\mathscr{M}_d$
\emph{only at the boundary}.
\end{enumerate} 

Using Rehren's theorem as a guide, one can think of the regions

\begin{equation}\label{pcone1}
\alpha_z(\mathscr{W}_0)\,=\,\{x^\mu:(z,x^\mu)\in\mathscr{W}_0\},\;
0<z<1
\end{equation}

as images of $\mathscr{W}_0$ under a family of maps $\alpha_z$ that
interwine the actions of the Poincar\'e (sub)group with respect to each 
leaf, just as $\alpha$ interwines the actions of the full $AdS$
group. Furthermore, by assigning to each region in (\ref{pcone1}) the
C*-algebra $\mathfrak{C}_z(\alpha_z(\mathscr{W}_0))\,:=\,\mathfrak{A}
(\mathscr{W}_0)$ and extending the nets to all $\mathscr{M}_d$ by 
Poincar\'e covariance, we get a family of (isotonous, Poincar\'e 
covariant) \emph{leaf nets} of observables, indexed by
$z$. The leaf nets possess, relatively to the boundary net, the 
following properties:

\begin{itemize}
\item\textbf{Scaling dynamics:}\quad$\forall\,0<\lambda\leq 1,\,
0<z<1,$
\begin{equation}\label{rg1}
\mathfrak{B}(\lambda\alpha(\mathscr{W}_0))\,=\,\mathfrak{C}_{\lambda
z}(\lambda\alpha_{\lambda z}(\mathscr{W}_0))\end{equation} (remember 
that the dilations act covariantly on the
boundary net); and, as a natural consequence of this,
\item\textbf{Algebraic UV/IR connection:}\quad$\forall\,0<z<\lambda\leq 1,$
\begin{equation}\label{rg2}
\mathfrak{B}(\lambda\alpha(\mathscr{W}_0))\,=\,\mathfrak{C}_z(
\lambda\alpha_z(\mathscr{W}_0)).
\end{equation}
\end{itemize}

The last property requires some explanation. The physical content of
the leaf nets is merely a description of the relation between the 
energy-momentum behaviour of the bulk and the boundary theories. To show this, 
we'll perform the construction of the respective \emph{scaling algebras} over
\emph{Minkowski} spacetime, following the work of Buchholz and 
Verch\cite{buch1}. 

\textsc{Definition\rm\cite{buch1}.}\quad\emph{Let
$\{\mathfrak{A}(\mathscr{O})\}$ be a net of local observables in
$\mathscr{M}_d$. The corresponding {\bf scaling net} is given by
assiging to each region $\mathscr{O}\subset\mathscr{M}_d$ the
{\bf scaling algebra} $\underline{\mathfrak{A}}(\mathscr{O})$ 
of (uniformly continuous and bounded) functions $\underline{A}:\lambda
\in(0,a)\mapsto\underline{A}_\lambda\in\mathfrak{A}(\lambda
\mathscr{O})$ ($a$ can be infinite), such that:
\begin{enumerate}
\item $\underline{\mathfrak{A}}(\mathscr{O})$ is a C*-algebra under
pointwise C*-algebraic operations, with the C*-norm $\|\underline{A}\|
\,=\,\sup_{\lambda\in(0,a)}\|\underline{A}_\lambda\|$;
\item The action $\delta_{\Lambda,x}$ of the Poincar\'e group on the
underlying net can be lifted to the scaling net by taking 
$(\underline{\delta}_{\Lambda,x}(\underline{A}))_\lambda\,=\,
\delta_{\Lambda,\lambda x}(\underline{A}_\lambda)$; we assume that 
it's norm continuous: $\|\underline{\delta}_{\Lambda,x}(\underline{A})
-\underline{A}\|\,\stackrel{(\Lambda,x)\rightarrow
(\mathbf{1},0)}{\longrightarrow}\,0,\,\forall\underline{A}$;
\item Scaling transformations act in a covariant and norm continuous
manner, through $(\underline{\sigma}_\mu(\underline{A}))_\lambda\,:=\,
\underline{A}_{\mu\lambda},\,\forall\mu:\mu\lambda<a$.
\end{enumerate} 
}

Physically, the scaling net corresponds to all admissible (multiplicative)
renormalization prescriptions in the underlying theory, which, of
course, must be all physically equivalent if the theory has
sufficiently good UV behaviour. An important question to be asked is: 
\emph{how do the scaling algebras $\underline{\mathfrak{B}}(\alpha
(\mathscr{W}_0))$ and $\underline{\mathfrak{C}_z}(\alpha_z(
\mathscr{W}_0))$ compare?} The answer is simple and is given by 
(\ref{rg2}): the scaling properties coincide up to the scale 
$\lambda=z$, for each corresponding leaf net. Should we ignore 
the contribution of the scaled observables below $\lambda=z$ (UV 
cutoff), the corresponding leaf net will behave just like the 
boundary theory dual to a ``IR-cutoff'' $AdS$ net. This somewhat 
loose explanation suggests the name ``algebraic UV/IR 
connection'', since a similar phenomenon occurs in the (stringy)
AdS/CFT correspondence\cite{suss1}. 

\section*{Acknowledgements} 

This is a slightly modified version of a poster presentation made at
this Conference. The author would like to thank Kostas Skenderis for
the enlightening discussions and the subsequent email exchange, and
Karl-Henning Rehren for pointing out an error in a previous version.

\end{document}